\documentclass[journal,letterpaper,twoside,twocolumn]{IEEEtran}
\usepackage{amsmath,amssymb,graphicx,psfrag,cite}
\usepackage[normalem]{ulem} 
\usepackage[usenames,dvipsnames]{color} % red, green, Green, Tan, SkyBlue...
\usepackage{balance}
\usepackage{mathrsfs}   % Define \mathscr, a script font
\usepackage{paralist}

\graphicspath{{figure/}}

\usepackage{color}

\title{Performance Metrics for Systems with Soft-Decision FEC and Probabilistic Shaping}
\author{Tsuyoshi~Yoshida,~\IEEEmembership{Member,~IEEE,} Magnus~Karlsson,~\IEEEmembership{Senior~Member,~IEEE,} \\ and~Erik~Agrell,~\IEEEmembership{Senior~Member,~IEEE} \thanks{Manuscript received April 28, 2017; revised June 12, 2017; revised Aug. 31, 2017; revised Oct. 2, 2017.}}%
\begin{document}
\maketitle

\begin{abstract}
High-throughput optical communication systems utilize binary soft-decision forward error correction (SD-FEC) with bit interleaving over the bit channels. The generalized mutual information (GMI) is an achievable information rate (AIR) in such systems and is known to be a good predictor of the bit error rate after SD-FEC decoding (post-FEC BER) for uniform signaling. However, for probabilistically shaped (nonuniform) signaling, we find that the normalized AIR, defined as the AIR divided by the signal entropy, is less correlated with the post-FEC BER. 
We show that the information quantity based on the distribution of the single bit signal, and its asymmetric log-likelihood ratio, are better predictors of the post-FEC BER.
In simulations over the Gaussian channel, we find that the prediction accuracy, quantified as the peak-to-peak deviation of the post-FEC BER within a set of different modulation formats and distributions, can be improved more than $10$ times compared with the normalized AIR.
\end{abstract}

\begin{IEEEkeywords} 
	Bit error rate, bitwise decoding, forward error correction, generalized mutual information, modulation, mutual information, optical fiber communication, probabilistic shaping.
\end{IEEEkeywords}

\vspace{-0.2cm}
\section{Introduction}
\label{sec:intro}
Performance metrics are key in the design, evaluation, and comparison of communication schemes. In optical systems, the bit error rate (BER) of the received data is traditionally the most important metric to quantify the performance. In modern optical communications, the BER requirement is typically $<10^{-15}$. 
After forward error correction (FEC) was introduced, it became desirable to find other performance metrics than the received BER after FEC decoding, the so-called \emph{post-FEC BER}. This is because measurements or simulations of very low BER after FEC decoding are time-consuming and of less general significance as they only apply to the chosen FEC code. In addition, FEC decoder hardware is usually not available in most laboratories.

For this purpose, the BER before the decoder, the \emph{pre-FEC BER}, or the Q-factor derived from this BER value, became common to predict the post-FEC BER \cite{ITU00}. These metrics work reasonably well with hard-decision FEC decoding and binary modulation in each dimension, such as on--off keying, binary phase-shift keying, or quaternary phase-shift keying.

The deployment of coherent detection with digital signal processing \cite{roberts_2009} made modulation formats more complex and diverse, and soft-decision (SD) FEC became widely utilized \cite{chang_2010_cm}.  In such systems, the pre-FEC BER is insufficient to predict the post-FEC BER, especially for the purpose of comparing different modulation formats. The mutual information (MI) as a metric was introduced in optical communications in \cite{leven_2011} and was considered for other channels in, e.g., \cite{wan_2006,franceschini_2006}. The MI is well suited for optical systems using coded modulation with nonbinary FEC codes or bit-interleaved coded modulation (BICM) with iterative demapping \cite{schmalen_2017}. However, even if such schemes work well in theory, practical optical systems often use bitwise receivers, i.e., BICM without iterative demapping, due to the simpler implementation, 
and the MI does not accurately predict the post-FEC BER of BICM schemes \cite{alvarado_2015}.
Instead, Alvarado \emph{et al.} proposed using the 
generalized MI (GMI), which is an achievable information rate (AIR), 
as a performance metric in optical communications with binary SD-FEC and bitwise decoding \cite{alvarado_2015}.

Shaping is now receiving wide interest in the optical communications community, as a means to close the gap between AIRs with common uniform quadrature amplitude modulation (QAM) formats and the capacity-achieving nonuniform input distribution \cite{bocherer_2015,bocherer_2017,shulte_2016,buchali_2016,fehenberger_2016,cho_2016,cho_2017}. 
In this paper, we investigate, for the first time, how well certain information-theoretic metrics can predict the post-FEC BER in probabilistically shaped systems with binary FEC codes and bitwise SD decoding. It turns out that the normalized AIR with the signal entropy 
%\uline{
is not as good metric in this case as with uniform signaling. %}
A metric recently proposed in \cite{cho_2017} will be discussed as well. 
Numerical simulations show that two other metrics, based on the distribution of the single bit stream and the conditional distribution of log-likelihood ratios (LLRs or L-values), have significantly better correlation with the post-FEC BER.

\vspace{-0.2cm}
\section{System model and performance metrics}
In this section, the system model is described and the considered performance metrics are introduced.
\label{sec:system}

\vspace{-0.2cm}
\subsection{System model with nonuniform signaling}
\begin{figure}[t]
	\begin{center}
		\setlength{\unitlength}{.6mm} %
		\scriptsize
		\includegraphics[scale=0.1]{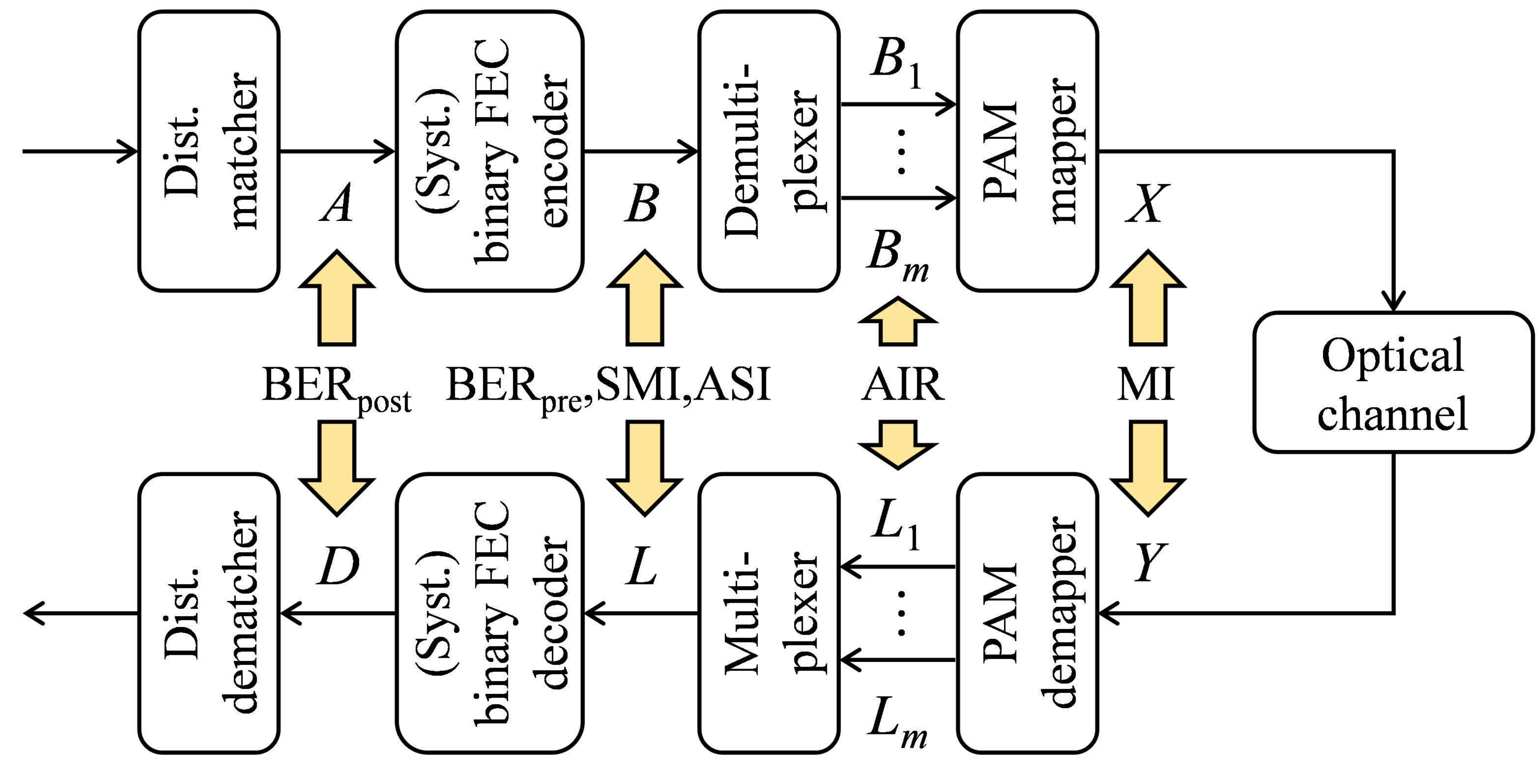}\\
		\vspace{-0.3cm}
		\caption{System model for nonuniform PAM signaling with bit-interleaved binary SD-FEC.}
		\label{fig:system}
	\end{center}
	\vspace{-0.6cm}
\end{figure}

Fig.~\ref{fig:system} shows the system model under consideration, which is the state-of-the-art configuration described in \cite{bocherer_2015}. The source data stream, consisting of independent, uniform bits, is processed in a distribution matcher and converted into a nonuniform bit sequence $A$. The binary FEC encoder treats $A$ as information bits and creates the bit sequence $B$ by inserting parity bits. When nonuniform (probabilistically shaped) signaling is used, the FEC encoding has to be systematic. 
Interleaving can be included in the encoder, but the nonuniform distribution must not be changed by encoding or interleaving. The bit sequence $B$ is demultiplexed into $m$ parallel bit tributaries (bit channels) $B_1, B_2, \cdots , B_m$ used for $2^m$-PAM mapping to the symbol $X$. 
The received PAM symbol $Y$ from the optical channel is demapped to the bit tributaries' LLRs $L_1,L_2,\cdots , L_m$.
These are multiplexed into a single LLR sequence $L$ and decoded. The decoder jointly processes all bit tributaries. The decoded bit sequence $D$ is finally recovered by the distribution dematcher. Bit errors have to be eliminated before distribution dematching to avoid error propagation.
This system model can apply also to uniform signaling
by removing the distribution matcher and the dematcher. 

\vspace{-0.2cm}
\subsection{Performance metrics for uniform signaling}
The BER after FEC decoding, $\text{BER}_{\text{post}}$, and before decoding, $\text{BER}_{\text{pre}}$, are defined as, resp.,
\begin{IEEEeqnarray}{rCL}
	\text{BER}_{\text{post}} &=& \sum_{b\in \{ 0,1 \}} P_{A,D}(b,1-b), \\
	\text{BER}_{\text{pre}} &=& \sum_{b\in \{ 0,1 \}} P_{B,\text{sign}(L)}(b,(-1)^{1-b}),
\end{IEEEeqnarray}
where $P$ denotes the joint probability of the indicated random variables.
The normalized GMI, which is a GMI (AIR with uniform signaling) normalized by the sent rate, is defined via the mutual information $I(B_i;L_i)$ between $B_i$ and $L_i$ as 
\begin{IEEEeqnarray}{rCL}
	\label{eq:GMI_per_m}
	{I}_{\text{n}} &=& \frac{1}{m} \sum_{i=1}^{m} I(B_i;L_i) .
\end{IEEEeqnarray}
where $L_i$ is the \emph{a posteriori} LLR \cite[eq.~(3.31)]{bicmbook}, defined as
	$L_i = \log (L_{i,0}/L_{i,1})$, $L_{i,b}=\sum_{\boldsymbol{b}\in \{0,1 \}^m: b_i=b} P_{\boldsymbol{B}}(\boldsymbol{b}) p_{Y \mid \boldsymbol{B}} (y \mid \boldsymbol{b})$,
$\boldsymbol{B} = [B_1,B_2,\cdots B_m ]$, and $p_{Y \mid \boldsymbol{B}}(y \mid \boldsymbol{b})$ is the real-valued 
channel assumed in the demapper \cite[eq.~(9)--(11)]{buchali_2016}.
To predict $\text{BER}_{\text{post}}$ in the case of uniform signaling with a binary FEC and bitwise SD decoding, $I_{\text{n}}$ is better than $\text{BER}_{\text{pre}}$ \cite{alvarado_2015}. 

In this paper, the receiver assumes a Gaussian $p_{Y \mid \boldsymbol{B}}$. 
In this case, the AIR of a uniform signal set can be estimated from the ensemble of transmitted bits $B_i(j)$ at time index $j$ and the corresponding LLRs $L_i(j)$ over $n_{\text{s}}$ symbols as \cite[eq.~(30)]{alvarado_2015}
\begin{IEEEeqnarray}{rCL}
	\label{eq:MCI}
	%I_{\text{n}} \approx \hat{I}_{\text{n}} &=& 1 - \frac{1}{m n_{\text{s}}}\sum_{i=1}^{m}\sum_{j=1}^{n_{\text{s}}} f(B_i(j),L_i(j)) , \\
	\hat{I} &=& 1 - \frac{1}{m n_{\text{s}}}\sum_{i=1}^{m}\sum_{j=1}^{n_{\text{s}}} f(B_i(j),L_i(j)) , \\
	\label{eq:loss_info}
	f(\beta,\lambda) &=& \log_2 ( 1+ \exp ( -(-1)^{\beta} \lambda )). 
\end{IEEEeqnarray}
Once ensemble representations of $B_i(j)$ and $L_i(j)$ are provided, 
these ensembles can be joined into ensemble representations of $B(k)$ and $L(k)$ simply by disregarding the channel index $i$ and using sample index $k$. The statistical relations between $B$, $L$, and their corresponding tributaries $B_i$ and $L_i$ are $P_{B}(b) = (1/m) \sum_{i=1}^{m}P_{B_i}(b)$, $p_{L}(l) = (1/m) \sum_{i=1}^{m}{p_{L_i}(l)}$, and $p_{B,L}(b,l) = (1/m)\sum_{i=1}^{m}{p_{B_i,L_i}(b,l)}$. This is because $I_{\text{n}} = I_{\text{s}}$ for uniform signaling \cite[Theorem 4]{ivanov16}, where
\begin{IEEEeqnarray}{rCL}
	\label{eq:Is}
	I_{\text{s}} = I(B;L)
\end{IEEEeqnarray}
is called \emph{single-bit MI (SMI)} in this paper.

\begin{figure*}[tb]
	\begin{center}
		\setlength{\unitlength}{.6mm} %
		\includegraphics[scale=0.1]{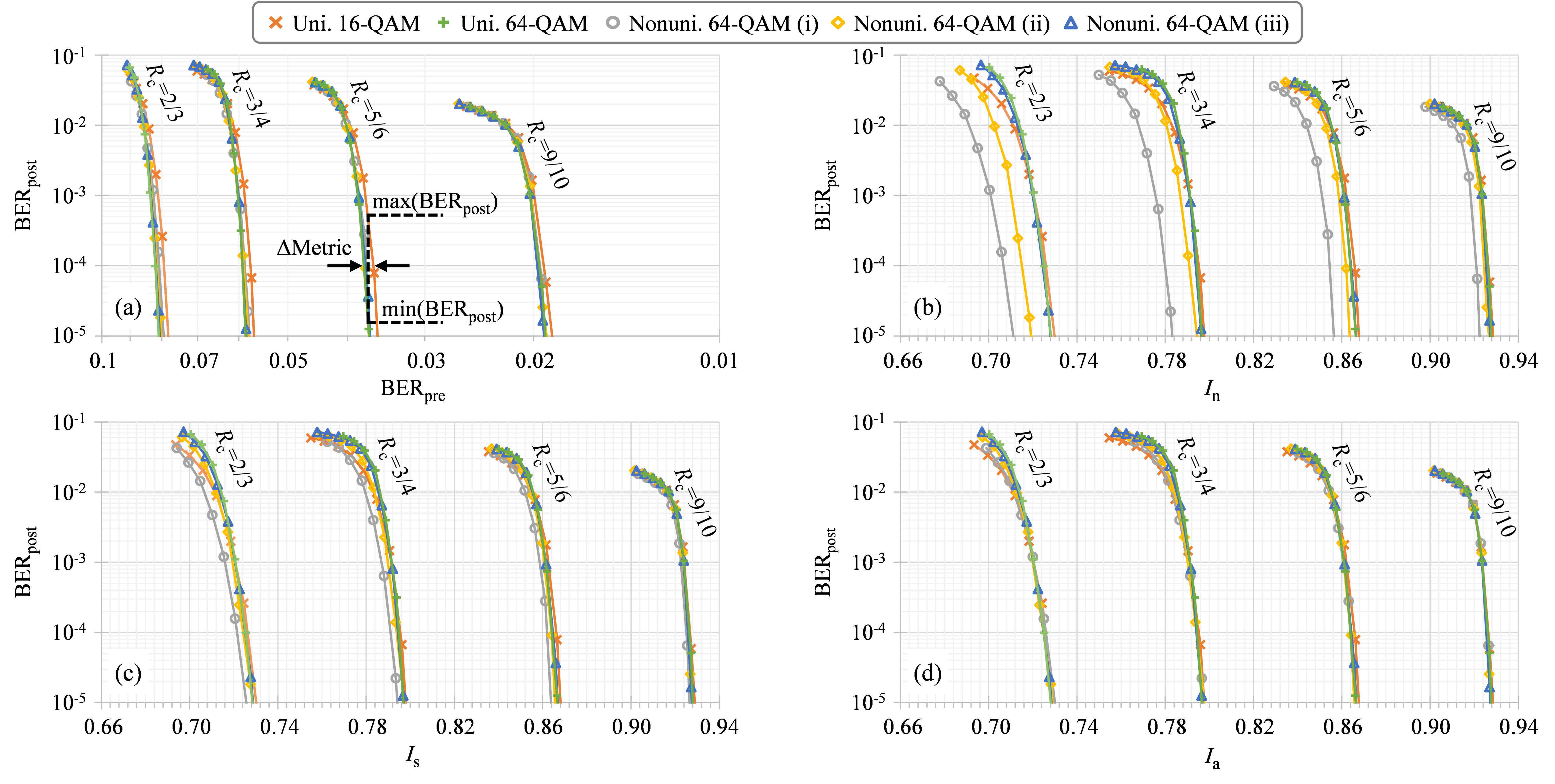} \\
		\vspace{-0.4cm}		
		\caption{Simulated post-FEC BER as a function of (a) pre-FEC BER $\text{BER}_{\text{pre}}$, (b) normalized AIR $I_{\text{n}}$, (c) SMI $I_{\text{s}}$, and (d) ASI $I_{\text{a}}$.}
		\label{fig:sim}
	\end{center}
	\vspace{-0.7cm}
\end{figure*}

\vspace{-0.2cm}
\subsection{Performance metrics for nonuniform signaling}
The metrics for nonuniform signaling, $I_{\text{n}}$ and $I_{\text{s}}$, need to be renormalized in order to span the range $[0,1]$. To this end, \eqref{eq:GMI_per_m} and \eqref{eq:Is} are replaced by the more general expressions
\begin{IEEEeqnarray}{rCL}
	\label{eq:In}
	I_{\text{n}} &=& R_{\text{BMD}}/H(\boldsymbol{B}) ,  \\
	\label{eq:SMI}
	I_{\text{s}} &=& I(B;L) / H(B), 
\end{IEEEeqnarray}
where $R_{\text{BMD}}$ in \eqref{eq:In} is an AIR for nonuniform signals \cite[eq.~(67)]{bocherer_2015} \cite[eq.~(45)]{bocherer_2017}, defined as
\begin{IEEEeqnarray}{rCL}
	\label{eq:RBMD}
	R_{\text{BMD}} = \max\left\{0, H(\boldsymbol{B}) -{\sum_{i=1}^{m} ( H(B_i)-I(B_i;L_i) )} \right\} ,
\end{IEEEeqnarray}
where $H(\boldsymbol{\mathcal{B}})=-\sum_{\boldsymbol{b}\in\{ 0,1 \} ^s} P_{\boldsymbol{\mathcal{B}}}(\boldsymbol{b})\log_2 P_{\boldsymbol{\mathcal{B}}}(\boldsymbol{b})$ is the entropy of the $s$-dimensional discrete variable $\boldsymbol{\mathcal{B}}$.

We further consider the information quantity from an SD-FEC decoder operation, which works based on the hard-decision value $\text{sign}(L)$ and the reliability $|L|$. If $|L|$ is large, there will be less probability to flip the hard-decision value. Thus, the \emph{asymmetric LLR} $L_{\text{a}} = (-1)^{B} L$,
which is included in \eqref{eq:loss_info}, is an important quantity, as it accounts for both the sign and the magnitude of the LLR.
The distribution
\begin{IEEEeqnarray}{rCL}
	\label{eq:PLv}
	p_{L_{\text{a}}}(l) &=& \sum_{b\in \{ 0,1 \} } P_B(b) p_{L \mid B}((-1)^b l \mid b), %, \\
\end{IEEEeqnarray}
where $p_{L \mid B}(l \mid b) = (1/P_B(b)) \sum_{i=1}^{m}{p_{B_i,L_i}(b,l)}$, should be as asymmetric as possible for better decoding performance.\footnote{For uniform signaling, the distribution \eqref{eq:PLv} can be obtained by setting $b=0$ in the distribution of ``symmetrized'' LLRs in \cite[eq.~(19)]{ivanov16}.}
Then, we can quantify the asymmetry of $L_{\text{a}}$ by the \emph{asymmetric information (ASI)}
\begin{IEEEeqnarray}{rCL}
	\label{eq:ASI}
	I_{\text{a}} = 1 - h(L_{\text{a}} \mid |L_{\text{a}}|)=1+h(|L_\text{a}|)-h(L_{\text{a}}),
\end{IEEEeqnarray}
which is defined in $[0,1]$,
where $h(\mathcal{A}) = -\int_{-\infty}^{\infty} p_{\mathcal{A}}(a) \log_2 p_{\mathcal{A}}(a)\text{d}a$ is the differential entropy of the continuous variable $\mathcal{A}$.

All metrics defined for nonuniform signaling can be applied to uniform signaling having symmetric distributions ($P_B(0)=P_B(1)=1/2$ and $p_{L \mid B}(l \mid 0)=p_{L \mid B}(-l \mid 1)$), because it is just a special case of nonuniform signaling. In this case, $\sum_{i=1}^{m}H(B_i) = m$, $H(B) = 1$, $h(L)=1+h(|L_{\text{a}}|)$, and $h(L \mid B)=h(L_{\text{a}})$. Therefore, 
\eqref{eq:SMI} and \eqref{eq:ASI} are identical and $I_{\text{n}} = I_{\text{s}} = I_{\text{a}}$.

\begin{table}[tb]
	\caption{8-PAM signal parameters for simulation.}
	\label{tab:sim}
	\vspace{-0.4cm}
	\begin{center}
		\setlength{\unitlength}{.6mm} %
		\begin{tabular}{ccccc}
			\hline \hline
			Condition & Uni. & Nonuni. (i) & Nonuni. (ii) & Nonuni. (iii) \\ \hline
			$\sum_{i=2}^{m}N_{B_i} / N_{\text{s}}$ & 2 & 1.788 & 1.934 & 1.979 \\
			$\sum_{i=1}^{m}H(B_i)$ & 3 & 2.841 & 2.960 & 2.996 \\
			$H(B)$ & 1 & 0.9831 & 0.9967 & 0.9997 \\
			$H(\boldsymbol{B})$ & 3 & 2.803 & 2.951 & 2.995 \\ \hline \hline			
		\end{tabular}
	\end{center}
	\vspace{-0.7cm}
\end{table}

\vspace{-0.2cm}
\section{Simulations}
\label{sec:sim}
Here we will compare the above performance metrics as indicators of the post-FEC BER.
We apply \emph{probabilistic amplitude shaping %\uline{
(PAS)} %} 
\cite{bocherer_2015}, which is the current state-of-the-art shaping scheme in optical communications \cite{buchali_2016}.
Three nonuniform distributions based on 64-QAM were simulated by applying probabilistic shaping to Gray-coded 8-PAM in each dimension. To generate these distributions, a constant-composite distribution matcher was employed \cite{shulte_2016} with output block length $N_{\text{s}} =1024$.\footnote{A smaller $N_{\text{s}}$, such as $N_{\text{s}}=12$ in \cite{cho_2016}, gives simple implementation but reduces the rate. The rate loss is limited to $<0.55\%$ in the cases tested here.}
The probabilities used are the same as in  \cite[Table~I (a)--(c)]{fehenberger_2016}, and some measures of their nonuniformity are indicated in Table~\ref{tab:sim}.
$B_1$ is the sign bit, carrying the uniform information bits of the PAM symbol or FEC parity bits, and $B_2,\cdots B_m$ are amplitude bits. The number of input bits of bit tributary $B_i$ to the distribution matcher per block is denoted by $N_{B_i}$.
As benchmarks, uniform signaling using 16-QAM and 64-QAM was also simulated. As SD-FEC codes, we utilized the DVB-S2 binary low-density parity check codes \cite{dvbs2}, having a codeword length of 64800.
The bit mapper was (3,2,1) for 8-PAM, which is optimum \cite[Tab. V]{bocherer_2015}, and (2,1) for 4-PAM.
The examined code rates $R_{\text{c}}$ were $2/3$, $3/4$, $5/6$, and $9/10$. The results were averaged over 500 codewords in each case, and a couple of independent simulations were performed to verify that 500 codewords is indeed sufficient for reliable statistics.
The optical channel was assumed to be Gaussian with the signal-to-noise ratio varied with 0.1 dB granularity.
The FEC decoder's internal calculations used floating-point precision with 20 decoder iterations.

The metrics $I_{\text{n}}$, $I_{\text{s}}$, and $I_{\text{a}}$ were estimated from \eqref{eq:In}, \eqref{eq:SMI}, and \eqref{eq:ASI}, resp. The corresponding differential entropies were calculated as in \cite[Sec.~8.3]{cover06} from discretized versions of the LLRs $L_i$, $L$, and $L_{\text{a}}$, whose distributions were estimated using histograms having $2^5$ levels with optimized steps. The estimated error introduced by this discretization is only $<10^{-3}$ in each information quantity and hence negligible.

The results are shown in Fig.~\ref{fig:sim}, where it is desirable that all curves for the same code rate should be close to each other, if the quantity on the horizontal axis is to be useful to predict the post-FEC BER.
For most of the metrics, the difference due to modulation format and nonuniformity tends to increase at lower code rates $R_{\text{c}}$.
According to Fig.~\ref{fig:sim}(a), $\text{BER}_{\text{pre}}$ works surprisingly well for these signal sets
and will provide guidelines for real system performance.
Fig.~\ref{fig:sim}(b) shows that the normalized AIR $I_{\text{n}}$ is not a good metric for nonuniform signaling.
The required $I_{\text{n}}$ for nonuniform signaling is clearly less than that for uniform signaling to achieve the same $\text{BER}_{\text{post}}$ value such as $10^{-4}$. 
Fig.~\ref{fig:sim}(c) indicates that the SMI $I_{\text{s}}$ is better correlated with $\text{BER}_{\text{post}}$ for nonuniform signaling than $I_{\text{n}}$. This is a benefit of changing the treatment from parallel bit channels into a single bit sequence, which is what the FEC encoder and decoder will see.
Fig.~\ref{fig:sim}(d) shows that the ASI performs better than the other metrics by using the asymmetric distribution of $L_{\text{a}}$.
%\sout{Finally, we examined the metric $\mathit{NGMI} = (\sum_{i=1}^{m}I(B_i;L_i)+m-H(\boldsymbol{B}))/m$ proposed in} \cite{cho_2017}. 
%\sout{The results (not shown) indicate that the BER curves were spread significantly wider than with $I_{\text{s}}$ in the tested cases.}

Intuitively, the lower accuracy of $I_{\text{n}}$ can be understood because \eqref{eq:In} does not properly take FEC coding in the PAS scheme into account.
The maximum FEC code rate for error-free transmission is $R_{\text{c,max}}=1-(H(\boldsymbol{B})-R_{\text{BMD}})/m$, which is the code rate at which the information rate $H(\boldsymbol{B})-(1-R_{\text{c}})m$ equals the AIR $R_{\text{BMD}}$. $I_{\text{n}}$ tends to be lower than $R_{\text{c,max}}$ and the difference becomes larger for a strongly shaped signal.
%\sout{It should be noted that if the first term of $\mathit{NGMI}$ is replaced with $R_\text{BMD}$, then $R_{\text{c,max}}$ is obtained.}
%\uline{
The metric $\mathit{NGMI} = 1-(H(\boldsymbol{B}) - \mathit{GMI})/m$ %\sout{$(\mathit{GMI}+m-H(\boldsymbol{B}))/m$ proposed in} %} 
\cite{cho_2017}, 
%\uline{
where $\mathit{GMI}$ %for nonuniform signaling 
can be defined as $H(\boldsymbol{B})-\sum_{i=1}^{m} H(B_i \mid Y) = H(\boldsymbol{B})-\sum_{i=1}^{m} H(B_i \mid L_i)$,%}
\footnote{
	%\uline{
	%Though there is no such explicit definition to the best of our knowledge, 
	$H(\boldsymbol{B})$ minus the conditional entropy due to suboptimal reception is a natural extension of the GMI
	\cite[eq.~(21)]{alvarado_2015} to nonuniform signaling.
	Several kinds of rates in PAS are discussed in \cite{bocherer_2017}, too.
	}
%\uline{ 
 is identical to $R_{\text{c,max}}$ in the regime where $R_{\text{BMD}} > 0$. %}
On the other hand, %\sout{in the regime where $R_{\text{BMD}} > 0$,} 
one can show that %\sout{$R_{\text{c,max}}$} 
$I_{\text{a}}=\mathit{NGMI}$ by considering \cite[eqs.~(3.67), (4.81)]{bicmbook} and \cite[Sec. VI-B]{bocherer_2015}.
%\uline{
Note that \eqref{eq:MCI} represents the estimates of $\mathit{NGMI}$ and $I_{\text{a}}$ by Monte-Carlo integration for nonuniform signaling. %}

\begin{figure}[t]
	\begin{center}
		\setlength{\unitlength}{.6mm} %
		\scriptsize
		\includegraphics[scale=0.1]{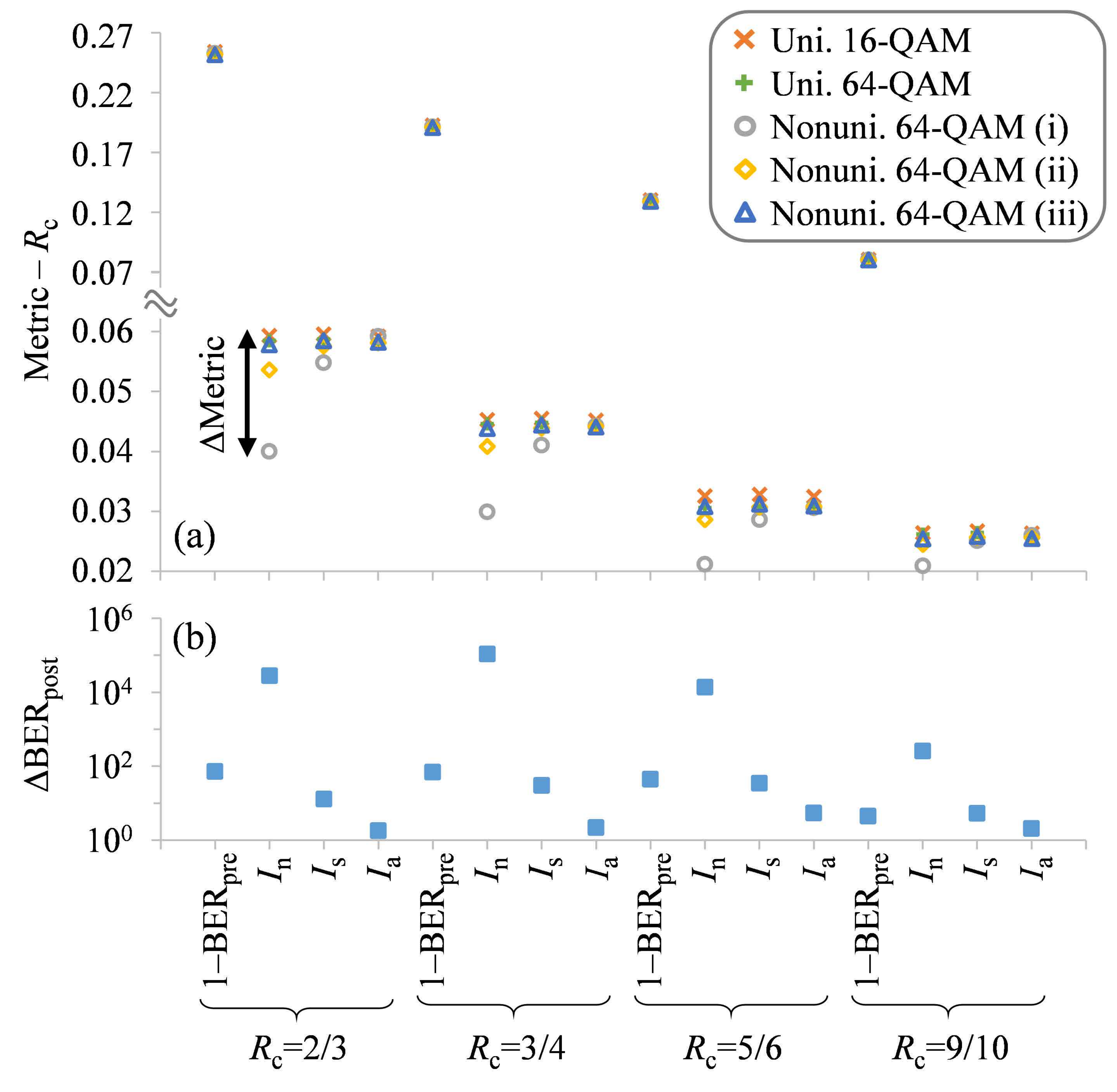}
		\vspace{-0.4cm}
		\caption{Comparison of prediction accuracy for (a) the metrics at $\text{BER}_{\text{post}}=10^{-4}$, and (b) %\sout{the post-FEC BER variation}
			$\Delta \text{BER}_{\text{post}}$ around $\text{BER}_{\text{post}}=10^{-4}$ among the studied cases.}
		\label{fig:sim2}
	\end{center}
	\vspace{-0.7cm}
\end{figure}

Fig.~\ref{fig:sim2} summarizes the prediction accuracy at $\text{BER}_{\text{post}}=10^{-4}$. Fig.~\ref{fig:sim2}(a) compares the information quantities $1-\text{BER}_{\text{pre}}$, $I_{\text{n}}$, $I_{\text{s}}$, and $I_{\text{a}}$. Relatively large deviations are mainly seen for nonuniform signaling (especially case (i), which is the most nonuniform) with the AIR metric. When considering all signals at the same BER, the maximum metric variations $\Delta \text{Metric}$ (see Fig.~\ref{fig:sim}(a)) in Fig.~\ref{fig:sim2}(a) are %$2.6\cdot 10^{-3}$, $1.9\cdot 10^{-2}$, $4.8\cdot 10^{-3}$, $2.0\cdot 10^{-3}$ %\sout{
0.0026, 0.019, 0.0048, and 0.0020 %} 
for $1-\text{BER}_{\text{pre}}$, $I_{\text{n}}$, $I_{\text{s}}$, and $I_{\text{a}}$, resp.
Note that the prediction accuracy for $1-\text{BER}_{\text{pre}}$, 
$I_{\text{n}}$, and $I_{\text{s}}$ tends to decrease for low $R_{\text{c}}$ or strong shaping, whereas the accuracy for $I_{\text{a}}$ does not. 
In comparisons of coding, shaping, and modulation schemes, a spectral efficiency difference of several percent tends to be relevant \cite{buchali_2016}, so a prediction variation in $I_{\text{n}}$ (0.019 in this case) may potentially affect such a comparison.
The accuracy depends on not only $\Delta \text{Metric}$ but also the steepness of the relationship of the metric with $\text{BER}_{\text{post}}$, 
so in Fig.~\ref{fig:sim2}(b) we show $\Delta \text{BER}_{\text{post}}=\text{max}(\text{BER}_{\text{post}})/\text{min}(\text{BER}_{\text{post}})$
around the specific metric value that gives $\text{BER}_{\text{post}}=10^{-4}$ on average. 
Thus $\Delta \text{BER}_{\text{post}}$ is a measure of the correlation between $\text{BER}_\text{post}$ and the chosen metric. 
The maximum $\Delta \text{BER}_{\text{post}}$ is $10^{1.9}$, $10^{5.0}$, $10^{1.5}$, and $10^{0.73}$ for $\text{BER}_{\text{pre}}$, $I_{\text{n}}$, $I_{\text{s}}$, and $I_{\text{a}}$, resp. 
$I_{\text{a}}$ is clearly the best metric having more than 10 times better accuracy than $\text{BER}_\text{pre}$ and $I_{\text{n}}$.
We tested stronger shaping cases like in \cite{buchali_2016}, and again found that $\Delta\text{BER}_{\text{post}}$ was consistent around $10^{-4}$ for $\text{BER}_{\text{pre}}$ and $I_{\text{a}}$ and varied much more ($\ge 10^{9}$) for $I_{\text{n}}$ and $I_{\text{s}}$.

\vspace{-0.2cm}
\section{Conclusions}
\label{sec:cncl}
We compared performance metrics to predict the post-FEC BER with binary SD-FEC in systems with probabilistic shaping. The normalized AIR has a large prediction variation of 
$10^{5}$ around $\text{BER}_{\text{post}}=10^{-4}$ with a low-density parity check code for nonuniform signaling, because it does not properly take the FEC coding into account. 
By employing the SMI and ASI metrics, the variation can be reduced to $30$ and $5$, resp.
While the prediction variation of normalized AIR and SMI increases in the case of lower $R_{\text{c}}$ or stronger shaping, the variation by ASI does not seem to depend thereon. Thus, once we know the ASI limit for a certain FEC code, we can infer the practical achievable rate by only calculating the ASI without FEC decoding.

\vspace{-0.2cm}
\section*{Acknowledgment}
We thank the anonymous reviewers for insightful comments about the role of FEC coding in PAS and %\sout{suggesting the above-mentioned modified $\mathit{NGMI}$ metric} %\uline{
the GMI concept for nonuniform signaling. %}

\vspace{-0.2cm}

\end{document}